\documentclass[11pt]{article}
\usepackage{Blois,epsfig}

\def\be{\begin{equation}}
\def\ee{\end{equation}}
\def\bea{\begin{eqnarray}}
\def\eea{\end{eqnarray}}

\begin{document}
\vspace*{2cm}
\begin{center}
\Large{\textbf{XIth International Conference on\\ Elastic and Diffractive Scattering\\ Ch\^{a}teau de Blois, France, May 15 - 20, 2005}}
\end{center}

\vspace*{2cm}
\title{Diffractive J/$\psi$ production in Ultraperipheral AuAu Collisions at RHIC}

\author{ Sebastian N. White, for the PHENIX Collaboration}

\address{Department of Physics, Brookhaven National Laboratory\\
Upton, NY 11973 USA}

\maketitle

\section{Introduction}
During the last decade measurements at HERA of inclusive hadron and exclusive vector meson photoproduction have contributed significantly to our understanding of the proton's structure. In exclusive J/$\psi$ photoproduction the cross sections are calculated assuming that the photon first fluctuates to a $c\bar c$  quark pair and then scatters off the proton target by exchange of a colorless pair of gluons. The t-distribution of vector meson production then measures the square of the ``two gluon form factor" which was found to be significantly more compact than the proton size inferred from elastic scattering.
	
	It would be interesting to continue the HERA measurements to higher center of mass energies and to nuclear targets but neither option is now considered as a HERA upgrade.
		On the other hand the LHC, when it collides Heavy (primarily Pb) Ions beams will provide higher energy $\gamma p$ and $\gamma$Pb collisions with enough luminosity to push beyond HERA in a single 1 month run\cite{svw}.
		In the meantime the RHIC diffractive physics program with ion beams previews the opportunities at the LHC. Diffractive  production of $\rho^o$ and two photon production of $e^+e^-$ pairs was measured by STAR\cite{starrho}.
		During the 2004 Au-Au run with $\sqrt{s_{NN}}=200$ GeV PHENIX commissioned a trigger for J/$\psi\rightarrow e^+e^-$ and the high mass di-lepton continuum. We report results from analysis of these data below.
		
\section{Expected Cross Section}
Electromagnetic Interactions between nuclei can be calculated for Ultraperipheral collisions (UPC), in which the impact parameter $b\geq2R_{nucleus}$, using the equivalent photon approximation. In this approach,
originally due to Fermi, one nucleus sees the strong E field of the incoming nucleus as an equivalent spectrum of quasi-real photons. As long as $Q^2$ of the exchanged photons is restricted to $Q^2<1/R_{nucleus}^2$ the spectrum is dominated by coherent emission from the nucleus with a large enhancement ($Z^2$) in the equivalent $\gamma$Nucleus luminosity. Because the strength parameter of the interaction, $\eta\sim Z^2\alpha$ , is close to unity additional low energy photon exchanges occur with high probability, particularly at small impact parameters\cite{Baur}. These few MeV photons are very effective in causing 1 or 2 neutron emission from the otherwise intact nucleus. The neutron tagging aspect of these interactions is particularly useful for triggering of UPC events and, when combined with a
rapidity gap requirement (along the beam emitting the photon) makes possible a powerful trigger selection in heavy ion collisions. The calculated fraction of J/$\psi$ events with at least one neutron tag is $60\%$.
	Several calculations can be found in the literature\cite{starlight} and an event generator, ``STARLIGHT" ,is used at RHIC to simulate both vector meson coherent production and $\gamma\gamma\rightarrow e^+e^-$. Recently\cite{Strikman} a calculation of incoherent J/$\psi$ production was also presented. This calculation considers the same coherently produced photon flux but instead of quasielastic J/$\psi $ production off the entire nucleus it considers the corresponding production off nucleons in the target. Signatures of the incoherent process are a broader $p_T$ distribution of produced J/$\psi$'s and a higher neutron multiplicity due to interactions of the recoiling nucleon within the nucleus.
\section{Trigger Selection}
	The PHENIX experiment has excellent capability for identifying electrons since it includes a high resolution electromagnetic calorimeter(EMC) array and Ring Imaging Cerenekov(RICH) counters. The RICH and EMC cover the same angular region ($|\eta|\leq$0.35) as the PHENIX tracking system in 2 approximately back-to-back spectrometer arms. In addition to the tracking coverage near $\eta$=0 a hodoscope array (BBC) covering 3.0$\leq |\eta_{BBC}|\leq$3.9 at both +z and -z is used for triggering on inelastic heavy ion collisions.  In the very forward direction Zero Degree Calorimeters (ZDC's) measure the number of neutrons from beam dissociation and can be used to trigger on 1 or more neutrons in either beam direction. 
		The PHENIX ultraperipheral di-electron trigger combined 3 of the above elements to select 1 or more beam dissociation neutrons, at least 1 electromagnetic cluster in the EMC and a rapidity gap signalled by no hits in one or the other of the BBC counters.
		
\begin{equation}
Trigger(UPC)=( EMC \geq 0.8 GeV)(ZDC(N)||ZDC(S) ) (\overline{BBC})
\end{equation}
	This very loose trigger yielded $8.5\times 10^6$ events while $1.12\times 10^9$ minimum bias interactions were recorded. So the Ultraperipheral trigger comprised less than $0.5 \%$ of the inelastic cross section and a negligible part of the available trigger bandwidth.
		
\begin{figure}[tbh]
\begin{center}
\includegraphics[width=10cm]{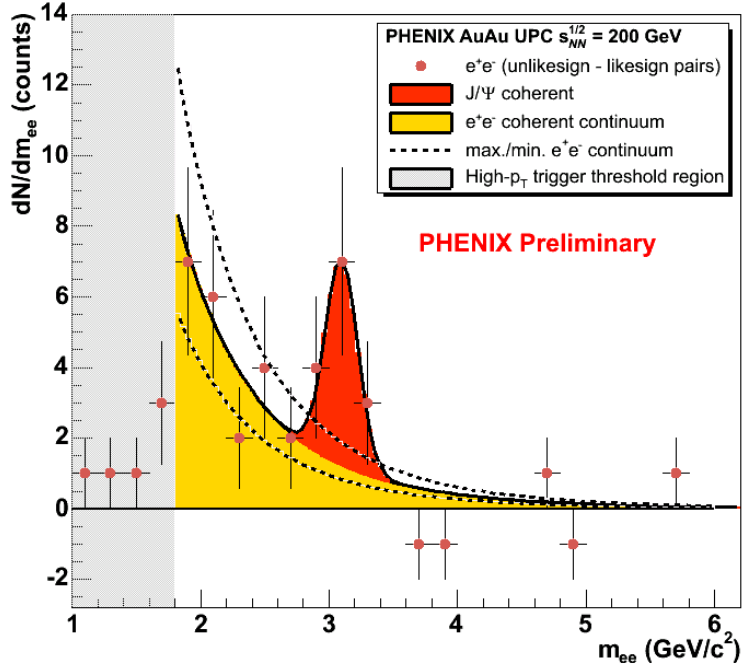}
\caption{Dilepton mass spectrum with dotted curve showing the range of continuum subtraction applied.}
\end{center}
\label{1}
\end{figure}

\section{Event Selection}

	The main features expected for photoproduced dileptons are small net transverse momentum of the pair and low multiplicity of produced tracks (both characteristic of diffractive processes). J/$\psi$'s are produced 	with y$\leq$1 .
		The first event selection we make is a charged track multiplicity in the PHENIX central detector of $n_{track}\leq15$. Then using tracks reconstructed in the PHENIX pad chamber we require that the reconstructed vertex fall within $|z_{vertex}|\leq 30\rm{cms}$.
		
			With the vertex cut applied and an additional removal of 21$\%$ of the data having different running conditions we calculate the integrated luminosity corresponding to our data sample. From the number of minimum bias interaction triggers in the remaining sample and the previously determined cross section $\sigma_{min.bias}=6.3\pm0.5$ barns\cite{ppg014} we find $L_{int}=120\pm10 \mu barn^{-1}$.

	The momentum of electron candidate tracks was measured using the deflection in the magnetic spectrometer. 	Having defined electron candidate trajectories and momenta in the spectrometers we then require consistent response in the RICH and EMCAl:
	\begin{itemize}
	\item{at least 2 photomultipliers have a cerenkov signal in the correct region of the RICH}
	\item{at least one electron deposits EMCAL energy greater than 1 GeV}.
	\end{itemize}

	Finally we make the simple requirement that electron candidates occupy different spectrometer arms since low pt J/psi events emit decay electrons back-to-back.
\begin{figure}[tbh]
\begin{center}
\includegraphics[width=8cm]{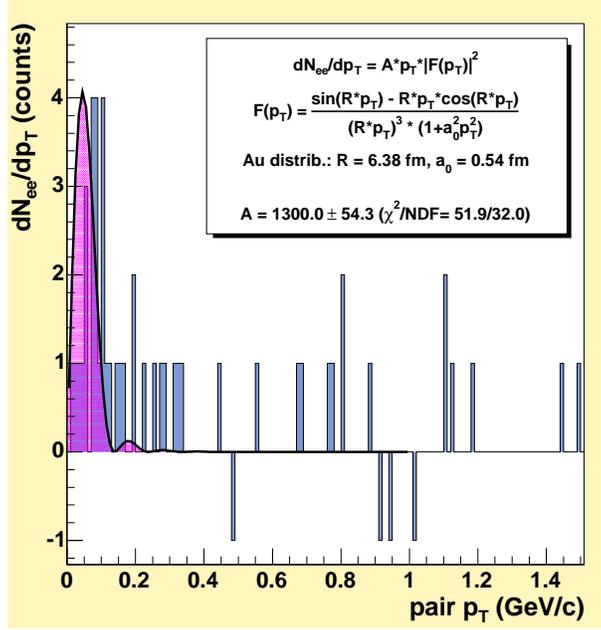}
\caption{Transverse momentum distribution of $e^+e^-$ pairs. The like sign pair distribution has been subtracted.}
\end{center}
\label{1}
\end{figure}

\section{Results}

	With the above cuts we find 42 oppositely charged electron candidates and 7 same charge with $m_{ee}\geq 1.8$ GeV/$c^2$. To estimate the combinatorial background we simply subtract the same sign candidate distribution from the opposite sign and obtain the spectrum shown in Fig.1. 
	
	To extract the J/$\psi$ signal we fit the reconstructed continuum spectrum from simulation to a power law and use the number of events in the 1.8-2.0 GeV bin to determine the size of the continuum subtraction and shown in Fig.1. The extracted signal is $10\pm 3(stat)\pm 3(syst) $ events with the continuum subtraction dominating the systematic error.
	
	Inclusive J/$\psi$ production in heavy ion collisions has a broad $p_T$ distribution (with an average of about 1.5 GeV/c)\cite{leitch}. This should be compared to the $p_T$ distribution shown in Fig. 2 where we combine all pairs with $m_{ee}>1.8$ GeV/$c^2$. Also shown in Fig. 2 is the expected shape due to the Au target form factor.
	
	We calculate the J/$\psi$ photoproduction cross section , correcting for detector acceptance and cut efficiencies both obtained by analyzing simulated J/$\psi$'s with the expected $p_T$ distribution. For J/$\psi$'s with $|y|\leq 0.5$ the geometrical acceptance and cut efficiencies are $5.0\%$ and $56.4\%$, respectively. We find
\begin{equation}
\frac{d\sigma_{J/\psi-UPC}}{dy}\times BR(ee)=48\pm14(stat)\pm16(syst)\mu b
\end{equation}
	at y=0, in good agreement with the prediction of 58 $\mu$b from ref.\cite{starlight}.
		
	In the next run with AuAu collisions PHENIX will see a 10-fold increase in event yield making possible a detailed study of both coherent and quasielastic J/$\psi$ production. We will also commission a second trigger sensitive to the $\mu^+\mu^-$ decay mode at large rapidity where the quasielastic signal will dominate\cite{Strikman}. Nevertheless the present low statistics measurement clearly demonstrates the feasibility of low cross section diffraction with Heavy Ion beams at RHIC and the LHC.

\section*{Acknowledgments}
 This work was supported 
in part under DOE Contract number DE-AC02-98CH10886.

\end{document}